\documentclass{birkjour}
\usepackage{graphicx}
\usepackage{amsmath}
\usepackage{amssymb}

\newcommand{\ii}{\mathrm{i}} %imaginary unit
\newcommand{\ee}{\mathrm{e}} %exponent
\def\d{\mathbin{\cdot}} %dot product with negative space between letters
\def\w{\mathbin{\wedge}} %wedge product with negative space between letters

\newcommand{\e}[1]{\mathbf{e}_{#1}} % Clifford basis vectors in Cl(3,1)
\newcommand{\Ie}[1]{I\negthinspace\mathbf{e}_{#1}}%bivector in Cl(3,1)
\newcommand{\inv}[1]{\overline{#1}} %space inversion
\newcommand{\ba}{\mathbf{a}} %bold a
\newcommand{\bk}{\mathbf{k}}
\newcommand{\bp}{\mathbf{p}}
\newcommand{\bx}{\mathbf{x}}
\newcommand{\bbf}{\mathbf{f}} %bold f

\def\sR{\mathsf{R}} %sans serif Rotation
\def\sH{\mathsf{H}} %sans serif Hamiltonian
\def\sF{\mathsf{F}} %sans serif F
\def\sG{\mathsf{G}} %sans serif G
\def\cB{\mathcal{B}} %bivector

\begin{document}


\title[Calculation of quantum eigens]
 {Calculation of  quantum  eigens\\  with geometrical algebra rotors}

%----------Author 1
\author{A.~Dargys}

\address{%
Center for Physical Sciences and Technology,\br
Semiconductor Physics Institute,\br
A.~Go\v{s}tauto 11, LT-01108 Vilnius, Lithuania}

\email{adolfas.dargys@ftmc.lt}

%\thanks{This work was completed with the support of our \TeX-pert.}
%----------Author 2
\author{A.~Acus}
\address{Institute of Theoretical Physics and Astronomy,\br
Vilnius University,\br
A.~Go\v{s}tauto 12, LT-01108 Vilnius, Lithuania}
\email{arturas.acus@tfai.vu.lt}
\subjclass{Primary 15A18; Secondary 15A66}

\keywords{Geometric algebra, rotors, quantum mechanics,
eigenspinors}

\date{October 3, 2014}

\begin{abstract}
A practical computational method to find the eigenvalues and
eigenspinors of quantum mechanical Hamiltonian is presented. The
method is based on reduction of the eigenvalue equation to well known
geometrical algebra rotor equation and, therefore, allows to replace the usual $\det (H-E)=0$ quantization condition by much simple vector norm preserving requirement. In order to show how it works in practice a number of examples are worked out in \textit{Cl}$_{3,0}$ (monolayer graphene and spin in the quantum well) and in \textit{Cl}$_{3,1}$ (two
coupled two-level atoms and bilayer graphene) algebras.
\end{abstract}

\maketitle

%%%%%%%%%%%%%%%%%%%%%%%%%%%%%%%%
\section{Introduction\label{sec:1}}\enlargethispage{3pt}
The standard approach to quantum mechanical eigenvalue problems is
based on matrix representation of the operator defined in a complex Hilbert space and its subsequent diagonalization in a chosen basis. A modern approach to
quantum mechanics \cite{Hestenes66,Snygg97,Doran03,Perwass09} is
based on Clifford's geometric algebra (GA), where the Clifford
space filled with geometric objects such as oriented lines, planes
etc is used instead of abstract Hilbert space. Selection of a
particular GA to solve the physical problem, in contrast to
Hilbert space approach, automatically then takes into account Clifford space
properties such as dimension, signature and symmetry relations of
the physical space or spacetime.

Despite the advantages that GA offers to quantum mechanics, in practice the methods needed to solve concrete
problems  are not elaborated enough. So, for the
physicists and engineers the GA may appear a nice but exotic
theory only. In this report a practical method to calculate eigenvalues and eigenmultivectors
is proposed that is based on the reduction of the  multivector
eigenvalue equation to well known rotor equation. In particular the primary goal of this paper is to shown by fully elaborated examples how GA can be used to find eigenvalues and eigenmultivectors in 3D~Euclidean space with \textit{Cl}$_{3,0}$ and in relativistic Minkowski space with \textit{Cl}$_{3,1}$ algebra without explicitly involving Hamiltonian matrix diagonalization procedure. Both mentioned algebras are of paramount importance in physics. 
An overview on eigenvalue problems in the framework of Clifford analysis is
given in Ref.~\cite{Sprossig01}. Some examples of GA approach to
eigenblades can be found in books~\cite{Snygg97,Doran03,Perwass09,Hestenes99}.

In the Sec.~\ref{sec:2} we shall remind the pertinent properties
of rotors and methods used to find the eigens. In Sec.~\ref{sec:3} we demonstrate how the
proposed method works in practice where we solve four problems of
increasing complexity. Relations with matrix representation
and symbols are summarized in Appendix. The following fonts are
used to denote GA elements:  $a$ or $\ba$ -- vector, $\mathcal{A}$ --
bivector, $A$ -- multivector, $\mathsf{A}$ -- transformation.

%%%%%%%%%%%%%%%%%%%%%%%%%%%%%%%%%%%%%%%%%%%%%%%%%%%%%%%
\section{Rotors and eigens in GA\label{sec:2}}

\textit{Rotation of vector}. In GA the rotation of vector $a$ to
vector $c$ is described by a linear transformation
\begin{equation}\label{rotvec}
\sR(a)=Ra R^{-1}=c,
\end{equation}
where $R$ is the rotor that satisfies $R R^{-1}=R^{-1}R=1$,
$R^{-1}=\tilde{R}$, with tilde denoting the reversion operation.
Since   $c^2=a^2$ the inverse transformation is $a=\tilde{R}c R$.
Usually the rotor $R$ is constructed applying two reflections in planes
perpendicular to unit vectors $\hat{m}$ and $\hat{n}$,
\begin{equation}
R=\hat{m}\hat{n},\quad \tilde{R}=\hat{n}\hat{m},\quad
\hat{m}^2=\hat{n}^2=1.
\end{equation}
In this way defined rotor can be put into more convenient exponential form
\begin{equation}\label{rotmn}
R=\hat{m}\hat{n}=\hat{m}\d\hat{n}+\frac{\hat{m}\w\hat{n}}{|\hat{m}\w\hat{n}|}|\hat{m}\w\hat{n}|=
\cos\frac{\theta}{2}+\hat{\cB}\sin\frac{\theta}{2}=\ee^{\hat{\mathcal{B}}\,\theta/2},
\end{equation}
where $\theta$ is the angle of rotation in a plane made up of
$\hat{m}$ and $\hat{n}$, $\cos(\theta/2)=\hat{m}\,\d\,\hat{n}$ and
$\sin(\theta/2)=|\hat{m}\w\hat{n}|$.
$\hat{\cB}=\frac{\hat{m}\w\hat{n}}{|\hat{m}\w\hat{n}|}$ denotes the
unit bivector, $\hat{\cB}^2=-1$.

In practice, however, it is required to have a rotor which brings
the end of a unit radius-vector $\hat{a}$ to a new position
$\hat{b}$. If $\theta$ is the angle between $\hat{a}$ an $\hat{b}$
then the rotation plane can be written as the outer product
$\hat{a}\w\hat{b}=(\hat{a}\hat{b}-\hat{b}\hat{a})/2$ and angle
$\theta$ can be expressed through the inner product
$\hat{a}\cdot\hat{b}=(\hat{a}\hat{b}+\hat{b}\hat{a})/2=\cos\theta$.
The rotor in this case takes the following form
\begin{equation}\label{rotab1}
R=\ee^{\pm\frac{\hat{a}\w\hat{b}}{|\hat{a}\w\hat{b}|}\frac{\theta}{2}}=
\cos\frac{\theta}{2}\pm\frac{\hat{a}\w\hat{b}}{|\hat{a}\w\hat{b}|}\sin\frac{\theta}{2},
\end{equation}
The trigonometric functions of half angle,
$\cos(\theta/2)=2^{-1/2}(1+\cos\theta)^{1/2}$ and
$\sin(\theta/2)=2^{-1/2}(1-\cos\theta)^{1/2}$, appear in the
formulas \eqref{rotmn} and \eqref{rotab1}. If $\hat{a}$ and
$\hat{b}$ are mutually orthogonal, then the
equation~\eqref{rotab1} simplifies to
\begin{equation}\label{rotab2}
R = \ee^{\pm\hat{a}\hat{b}\,\pi/4}=
\big(1\pm\hat{a}\hat{b}\big)\big/\sqrt{2},\quad \hat{a}\bot\hat{b}.
\end{equation}
The rotor can also be put~\cite{Doran03} in the form
$R=(1+\hat{a}\hat{b})/(|\hat{a}+\hat{b}|)$, which reduces
to~\eqref{rotab1} after half angle formulas are applied.

It is important to note, that two successive rotations can be
always described by the composite rotation
$\sR_3(a)=\sR_2\big(\sR_1(a)\big)\equiv\sR_2\sR_1a$. However, the
result of product of two rotors is not necessarily a rotor. The
exception happens, in particular, when $\sR_1$ and $\sR_2 $ belong
to nonintersecting subspaces.

\textit{Rotation of blade and multivector}. The rotation of a
blade relies on the extension of the outermorphism from vectors to
multivectors~\cite{Hestenes84}.  Since the outermorphism preserves
the outer product and as a result the shape of geometric objects,
this transformation induces a more general  linear transformation
in the full Clifford space. For example, the rotation of the
bivector $a\w b$ gives new bivector $\mathsf{R}(a\w
b)=\mathsf{R}(a)\w\mathsf{R}(b)$ having the same properties. To
rotate general blade $A_r=a_1\w a_2\w\dots\w a_r$ one rotates all
the vectors $a_i$ in it
$\underline{\sR}(A_r)=\sR(a_1)\w\sR(a_2)\w\dots\w \sR(a_r)$ (the
underbar denotes the outermorphism), so the rotation formula for
blade $A_r$ has the same structure as for vector~\eqref{rotvec},
$\mathsf{R}(A_r)=R A_r\tilde{R}=B_r$.  In particular, the
 pseudoscalar is rotationally invariant $RIR^{-1}=I$.
The formula remains valid if the blade is replaced by any
multivector. General properties of rotation transformation $\sR$
of multivectors $M$ and $N$ are listed below.
\[\begin{array}{lll}
1.& \sR(\alpha M)=\alpha\sR(M)&\textrm{Preserves scalar multiplication} \\
2.& \sR(M+N)=\sR(M)+\sR(N)&\textrm{Preserves multivector addition} \\
3.& \sR(M\d N)=\sR(M)\d\sR(N)&\textrm{Preserves inner product} \\
4.& \sR(M\w N)=\sR(M)\w\sR(N)&\textrm{Preserves outer product} \\
5.& \sR(MN)=\sR(M)\sR(N)&\textrm{Preserves geometric product} \\
6.&\textrm{grade}\big(\sR(A_r)\big)=\textrm{grade}(A_r)&\textrm{Preserves grade $r$}\\
\end{array}\]
An important property of the outermorphism is that the
outermorphism of the product of two functions $\sF$ and $\sG$ is
the product of the outermorphisms,
\begin{equation}
\underline{\sF\sG}(a\w b\dots\w c)=\sF\sG(a)\w \sF\sG(b)\dots\w
\sF\sG(c).
\end{equation}

\textit{Blade eigens}. The outermorphism allows to
characterize the invariant subspaces under linear transformations.
In particular, it allows to generalize the concept of the
eigenblade~\cite{Hestenes84}. Indeed, if $\sF$ is a linear function then
the equation
\begin{equation}\label{eigen}
\underline{\sF}(A_r)=\lambda A_r
\end{equation}
is called the eigenblade equation or collineation transformation,
where blade $A_r$ of grade $r$ is called eigenblade and $\lambda$ is
a real eigenvalue. Frequently the outermorphism symbol (underbar)
is omitted.  For example, if $a$ and $b$ are vectors and function
$\sF$ satisfies $\sF(a)=b$ and $\sF(b)=a$, then by outermorphism
extension we have $\underline{\sF}(a\w b)=\sF(a)\w \sF(b)=b\w
a=-a\w b$, or $\sF(a\w b)=- a\w b$. Thus $a\w b$ is the
eigenbivector with eigenvalue $-1$.

If applied to pseudoscalar $I$ the outermorphism can be used to define
the determinant $\text{det}(\sF)$ of linear transformation $\sF$,
\begin{equation}\label{det}
\underline{\sF}(I)=\text{det}(\sF)I,
\end{equation}
without any need of matrices. For example, in
\textit{Cl}$_{3,0}$ the determinant of $\sF$ as follows from
\eqref{det}  is
 $\text{det}(\sF)=\big(\sF(\e{1})\w\sF(\e{2})\w
\sF(\e{3})\big)\big(\e{3}\w\e{2}\w\e{1}\big)=\text{scalar}$.

The eigenvalue equation~\eqref{eigen} can be rewritten in the form
\begin{equation}
(\sF-\lambda)(A_r)=0
\end{equation}
showing that the operator $(\sF-\lambda)$ is singular. Every
singular operator has a vanishing determinant:
\begin{equation}\label{deteq}
\det(\sF-\lambda)(A_r)=\big(\sF(I_r)-\lambda I_r\big)I_r^{-1}=0,
\end{equation}
where the last expression follows from~\eqref{det}, the $I_r$ being
$r$-grade pseudoscalar. This equation is called the secular
equation for~$\sF$, the solution of which with respect to
$\lambda$ gives the eigenvalues. For example, let $\sF$ be a
vector-valued function $\bbf(\bx)$ of vector $\bx$ in
\textit{Cl}$_{3,0}$. If $\bbf_i=\bbf(\e{i})$, then
\[\text{det}(\bbf-\lambda)(\bx)=(\bbf_1-\lambda\e{1})\w
(\bbf_2-\lambda\e{2})\w(\bbf_3-\lambda\e{3})I_3^{-1}=0,\] where
$I_3=\e{1}\w\e{2}\w\e{3}=\e{1}\e{2}\e{3}$ and $I_3^{-1}=-I_3$.
After expansion this equation reduces to scalar cubic equation
\[\lambda^3+\alpha_2\lambda^2+\alpha_1\lambda+\alpha_0=0 \]
where the scalar coefficients are
$\alpha_0=(\bbf_1\w\bbf_2\w\bbf_3)\d I_3$,
$\alpha_1=-(\bbf_1\w\bbf_2\w\e{3}+\bbf_1\w\e{2}\w\bbf_3+
\e{1}\w\bbf_2\w\bbf_3)\d I_3$, and
$\alpha_2=(\bbf_1\w\e{2}\w\e{3}+
\e{1}\w\bbf_2\w\e{3}+\e{1}\w\e{2}\w\bbf_3)\d I_3$. After the
eigenvalues have been determined, the eigenblade
equation~\eqref{eigen} becomes algebraic multivector equation with
respect to  $A_r$ which can be solved by GA methods. Example of
applications of this approach can be found, for example, in the
textbook~\cite{Hestenes99}. The described method, however, is not
a coordinate free, and is based on expansion of the
transformation in a particular frame.

\textit{Spinor eigens}. The eigens, however, in many cases could
be calculated in a compact way without any reference to
determinant equation~\eqref{deteq}. In particular, as we shall
see, this is easy to do if the eigenequation
\begin{equation}\label{eigspin}
\sH(\psi)=E\psi
\end{equation}
for spinor $\psi$ that consists of a sum of even-grade blades can
be transformed to an eigenequation for a corresponding rotor. In quantum
mechanics the transformation $\sH$ is called the Hamiltonian and
$E$ is the energy (scalar). The reformulation of eigen
problem~\eqref{eigspin} may be realized if one observes that in
many quantum mechanical problems there is a specific direction
called the quantization axis ($z$-axis). As we shall see this
allows to reduce quantum mechanical eigenmultivector equation to a
single rotor equation, where the quantization axis is rotated to
some final direction determined by the Hamiltonian of the problem.
It is important to note that, in contrast to Hilbert space
formulation where the quantization axis is hidden in the structure
of the Hamiltonian, in GA this axis appears
explicitly. In other words the quantization axis, represented below by $\e{3}$
basis vector, appears directly in the Hamiltonian function. The
proposed method is based on this quantum mechanical
Hamiltonian property and allows us to construct a single optimal
rotor expressed in coordinate-free way, which is very similar to
rotor~\eqref{rotab1} or~\eqref{rotab2}, where the vectors
$a=\e{3}$ and $b$ are given by
eigenequation~\eqref{eigspin}.

The realization of the described procedure is straightforward in case
of \textit{Cl}$_{3,0}$ algebra (Schr{\"o}dinger-Pauli equation). In
relativistic \textit{Cl}$_{3,1}$ and \textit{Cl}$_{1,3}$ algebras
the spinor of Dirac equation is the sum of eight terms: scalar,
six bivectors and pseudoscalar. The reduction of relativistic
spinors to rotors of 3D~Euclidean space can be accomplished
after splitting of $\psi$ into even $\psi_{+}$ and odd $\psi_{-}$
parts with the help of spatial inversion operator described in 
Appendix~\ref{append},
\begin{equation}
\psi=\psi_{+}+\psi_{-},\quad\inv{\psi}=\psi_{+}-\psi_{-}.
\end{equation}
As we shall see the eigenvalue equations
$\sH(\psi_{+}+\psi_{-})=(\psi_{+}+\psi_{-})E$ and
$\inv{\sH}(\psi_{+}+\psi_{-})=(\psi_{+}-\psi_{-})E$ in this case
 reduce to two multivector equations for even $\psi_{+}$
and odd $\psi_{-}$ (or $I\psi_{-}$) parts. Reduced
equations then can be interpreted as rotors which rotate quatization axis $\e{3}$ to the final direction determined by the Hamiltonian.

%%%%%%%%%%%%%%%%%%%%%%%%%%%%%%%%%%%%%%%%%%%%%%%%%%
\section{Examples from quantum mechanics\label{sec:3}}
\begin{figure}
\centering
\qquad a)\includegraphics[width=3.cm]{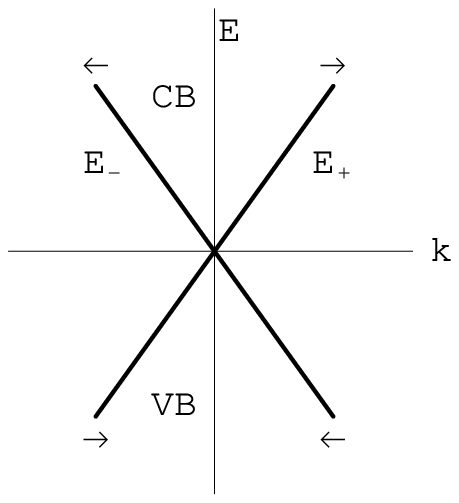}
b)\includegraphics[width=3.7cm]{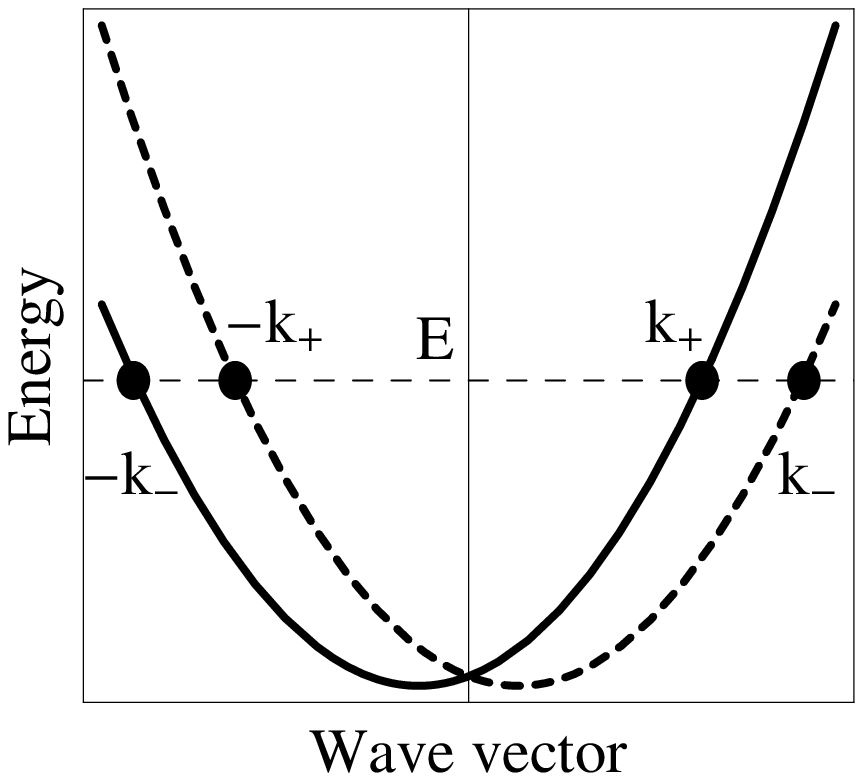}\newline
c)\includegraphics[width=4cm]{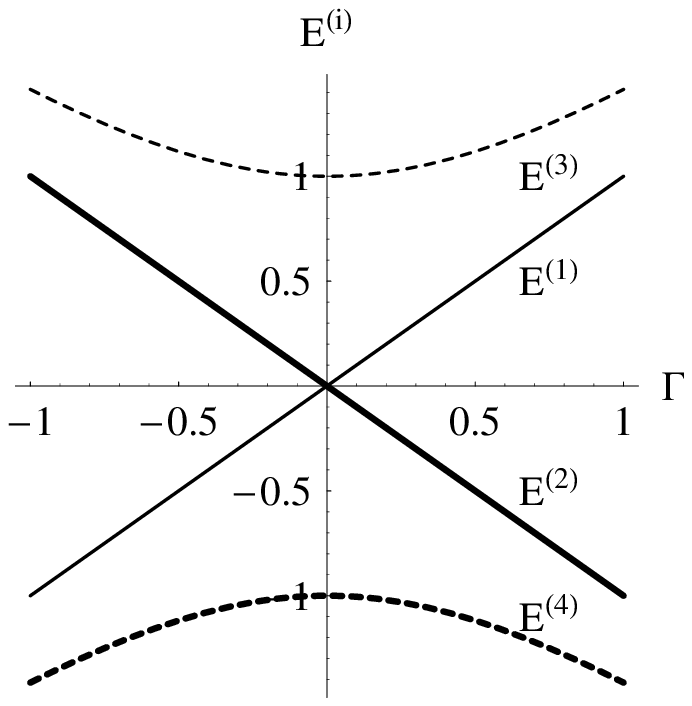}
d)\includegraphics[width=4cm]{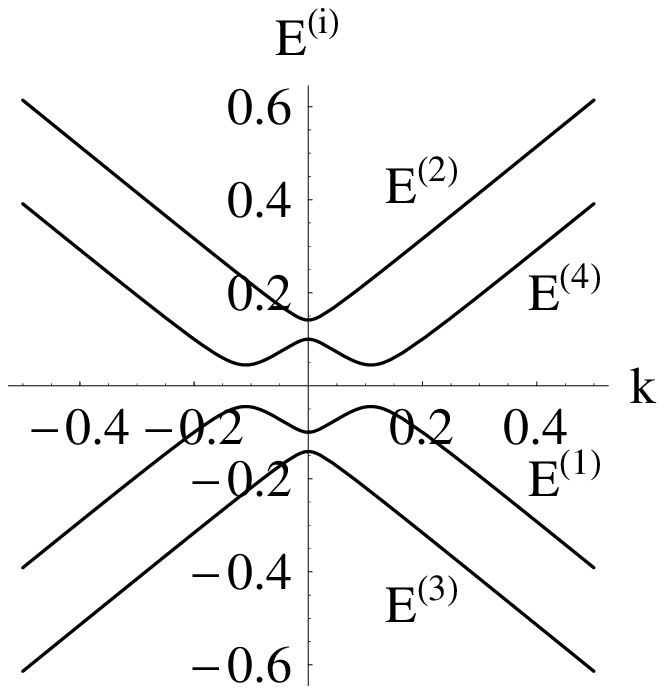}
\caption{a) Energy vs wave vector (spectrum) for a monolayer graphene. CB
 and VB are conduction and valence bands. The arrows show
 direction of pseudospin.
 b) Spectrum of the quantum well with spin-orbit interaction present. c) Energy of two coupled  two-level atoms vs coupling
constant $\Gamma$. d) Spectrum  of bilayer graphene with
interlayer voltage $2U$ applied.}\label{f:DispEx}
\end{figure}

In this section four characteristic examples from quantum
mechanics  are considered to illustrate how the proposed method
works in practice. The spectra of all considered Hamiltonians are
shown in Fig.~\ref{f:DispEx}.

%%%%%%%%%%%%%%%%%%%%%%%%%%%%%%%%%%%%%%%%%
\subsection{Monolayer graphene\label{sec:4a}}
Monolayer graphene is a 2D crystal consisting of a single layer of
carbon atoms. The intense interest in graphene was  stimulated by
2010 Nobel prize in physics (A.~Geim and K.~Novoselov) and
potential application of graphene in construction of novel 2D
nanodevices~\cite{Katsnelson12}. From theoretical point of view the most
interesting problem is the ultrarelativistic electron behavior in 2D
lattice at very small energies ($<1$~eV), Fig.~\ref{f:DispEx}a. The spectrum represents an ultrarelativistic limit of eigenenergy $E=\bigl(\bp^2 +m^{\star 2}\bigr)^{1/2}$ of the solution of the Dirac equation in the case when fraction of electron energy due to kinetic momentum $\bp$ is much larger than effective electron mass, $\bp^2 \gg m^{\star 2}$, where $m^{\star}\approx 0.012 m_0$ ($m_0$ being free electron mass).

\begin{figure}
 \centering
 \begin{minipage}[c]{0.35\textwidth}
 \raggedright
\includegraphics[width=5cm]{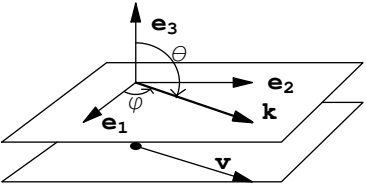}
 \end{minipage}%
 \begin{minipage}[c]{0.75\textwidth}
 \raggedleft
\caption{\label{f:grapheneAxes}2D electron (dot)  moving with
velocity $\mathbf{v}$ and coordinate system
$\{\e{1},\e{2},\e{3}\}$. The planes represent either two graphene
layers, or in the case of quantum well, two potential walls so that
the electron is squeezed between the planes. $\bk$ is the  wave
vector, $\theta=\pi/2$, $\cos\varphi=k_1/|\bk|$,
$\sin\varphi=k_2/|\bk|$.}
\end{minipage}%
\end{figure}

In the Hilbert space formalism the massless (ultrarelativistic)
Hamiltonian in the vicinity of critical $K$ point of graphene (the momentum space point where electron transport takes place) reduces to simple form
\begin{equation}\label{Hgmat}
\hat{H}_{MG}= \hbar v_F\begin{bmatrix}
 0& k_x-\ii k_y\\
 k_x+\ii k_y&0
\end{bmatrix},
\end{equation}
where $v_F\approx 10^{6}$~m/s is the Fermi velocity. The wave
vector $\bk=(k_x,k_y)$ lies in the graphene plane and is measured
with respect to $K$ valley minimum. In the following the natural
units $\hbar=1$ along with $v_F=1$ are used. The eigenenergies of
the Hamiltonian~\eqref{Hgmat} are
\begin{equation}\label{dispK2}
E_{\pm}=\pm\sqrt{k_x^2+k_y^2}=\pm|\bk|,
\end{equation}
where plus/minus sign  corresponds to conduction/valence bands
shown in Fig.~\ref{f:DispEx}a.  As discussed in~\cite{Dargys13a} a
minimal algebra needed  to describe the monolayer Hamiltonian is
\textit{Cl}$_{3,0}$, where the basis vectors $\e{i}$  play the
role of space coordinates and satisfy $\e{1}^2=\e{2}^2=\e{3}^2=1$,
Fig.~\ref{f:grapheneAxes}. The orthogonal oriented planes are
described by bivectors $\Ie{1}=\e{2}\e{3}$, $\Ie{2}=\e{3}\e{1}$,
and $\Ie{3}=\e{1}\e{2}$. With scalar and pseudoscalar
$I=\e{1}\e{2}\e{3}$ included, there are  8 basis elements.

Now we shall transform the Hamiltonian~\eqref{Hgmat} to
\textit{Cl}$_{3,0}$ algebra using the following mapping rules from
Hilbert to Clifford  space~\cite{Doran03},
\begin{equation}\label{repRules}
\begin{split}
|\psi\rangle &\longleftrightarrow\psi=a_0+a_1\Ie{1}+a_2\Ie{2}+a_3\Ie{3},\\
\lambda|\psi\rangle &\longleftrightarrow\lambda\psi,\quad
\hat{\sigma}_i|\psi\rangle \longleftrightarrow\e{i}\psi\e{3},
\end{split}
\end{equation}
where $\lambda$ is the scalar and $\hat{\sigma}_i$ is one of $2\times 2$ complex Pauli
matrices $\hat{\sigma}_1=\hat{\sigma}_x$, $\hat{\sigma}_2=\hat{\sigma}_y$ or $\hat{\sigma}_3=\hat{\sigma}_z$. Then, $\hat{H}_{MG}|\psi\rangle$ maps to the following
simple GA function of spinor $\psi$
\begin{equation}\label{HkinGA}
\sH_{MG}(\psi)=\bk\psi\e{3},
\end{equation}
where the wave vector $\bk=k_x\e{1}+k_y\e{2}$ lies in the graphene
plane, Fig.~\ref{f:grapheneAxes}. It should be noted that the
Hamiltonian function~\eqref{HkinGA} has a coordinate-free form.
The appearance of vector $\e{3}$ in Hamiltonian~\eqref{HkinGA} follows from the existence the pseudospin quantization axis, which is
perpendicular to graphene plane, Fig.~\ref{f:grapheneAxes}. This
is not so evident from the Hilbert space Hamiltonian~\eqref{Hgmat}.

The eigenvalue equation for monolayer graphene then reads
\begin{equation}
\bk\psi\e{3}=\psi E ,
\end{equation}
which can be easily rewritten in a form of the rotor equation
\begin{equation}\label{rotMG}
\frac{\psi\e{3}\tilde{\psi}}{\psi\tilde{\psi}}=\frac{\bk}{|\bk|^2}E,
\end{equation}
where $|\bk|\equiv k=\sqrt{\mathstrut{\vphantom{x^2}}\smash{k_x^2}+\smash{k_y^2}}$ is the magnitude of the
wave vector. Since~$\e{3}$ is a unit vector, it follows that after
rotation by $\psi$ the right hand side of \eqref{rotMG} must be
unit vector too. This means that the square of it satisfies
algebraic equation $(E/k)^2=1$, which in fact represents the quantization condition, and 
from which the dispersion law, $E_{\pm}=\pm k$, immediately follows.

The corresponding eigenmultivectors can be deduced from the same
equation~\eqref{rotMG} after insertion of $E_{\pm}$ values and
requiring the rotor to satisfy $\psi\tilde{\psi}=1$. Then, the
equation~\eqref{rotMG} reduces to
\begin{equation}\label{rot2}
\psi_{\pm}\e{3}\tilde{\psi}_{\pm}=\pm\hat{\bk},
\end{equation}
where $\hat{\bk}=\bk/|\bk|$ is the unit vector. Thus, according to
GA the eigenspinors $\psi_{\pm}$ of a monolayer graphene are
simply rotors that bring  the unit vector $\e{3}$ which is
parallel to the quantization axis to new direction $\pm\hat{\bk}$,
which is either parallel, as shown in Fig.~\ref{f:grapheneAxes},
or antiparallel to the electron wave vector $\bk$. No such
clear-cut geometrical interpretation for eigenspinors exists in
the Hilbert space. Using \eqref{dispK2}, solution of the rotor equation~\eqref{rot2} can
be written automatically as a rotation of $\e{3}$ in the unit
bivector plane $\hat{\bk}\w\e{3}$ by angle $\theta=\pi/2$ [see
Eq.~\eqref{rotab2}],%
\begin{equation}\label{psiMG}
\psi_{\pm}=\ee^{\pm\hat{\bk}\w\e{3}\,\pi/4}
=\frac{1}{\sqrt{2}}\Big(1\pm\hat{\bk}\w\e{3}\Big)=\frac{1}{\sqrt{2}}\Big(1\pm\hat{\bk}\e{3}\Big).
\end{equation}

It should be noted that the eigenspinors~\eqref{psiMG} are
coordinate-independent. The appearance of $\e{3}$ in~\eqref{psiMG}
is indispensable since the quantum mechanics requires to indicate
the direction of the pseudospin (or spin) quantization axis which
in our case was selected to be $\e{3}$. If, as shown in
Fig.~\ref{f:grapheneAxes},  angle $\varphi$ between $\e{1}$ and
$\bk$ is introduced the equation~\eqref{psiMG} can be written in
coordinate form as
$\psi_{\pm}=\big(1\pm\sin\varphi\Ie{1}\mp\cos\varphi\Ie{2}\big)/\sqrt{2}$.

From quantum mechanical point of view the pseudospin in graphene
has the same properties as spin. However, the spin is an intrinsic
property of electron, while the pseudospin is the property of
interacting $p$-orbitals of carbon atoms that make up the lattice.
Thus, the average of pseudospin\pagebreak vector may be found in exactly the
same way as for electron spin~\cite{Doran03},
\begin{equation}
\langle\mathbf{P}\rangle=\psi_{\pm}\e{3}\tilde{\psi}_{\pm}=\pm\bk/|\bk|=
\pm(\cos\varphi\,\e{1}+\sin\varphi\,\e{2}).
\end{equation}
We see, that the pseudospin is either parallel or antiparallel to
electron propagation direction, Fig.~\ref{f:DispEx}a. This allows to introduce the helicity
operator~\cite{Dargys13a}.

%%%%%%%%%%%%%%%%%%%%%%%%%%%%%%%%%%%%%%%%
\subsection{Electron in a quantum well (QW)\label{sec:4b}}
Let us now consider a potential well where the electron is
squeezed between two stepped-potential planes. The Hamiltonian for
electron in the  QW is
\begin{equation}\label{ham2D}
\hat{H}_{QW}=\big(\hbar^2/(2m^*)\big)(k_x^2+k_y^2)\hat{1}+\alpha_{\rm{R}}(k_y\hat{\sigma}_x-k_x\hat{\sigma}_y)
\end{equation}
where $m^*$ is the effective mass of electron, $k_x$ and $k_y$ are
the in-plane components and $\hat{1}$ denotes $2\times 2$ identity matrix. The first term is the kinetic energy and
the second term describes interaction of electron spin with its
orbital motion via spin-orbit interaction constant
$\alpha_{\rm{R}}$. For simplicity we shall assume that
$m^*=\hbar=1$. Using the replacement rules~\eqref{repRules} the
Hamiltonian~\eqref{ham2D} can be mapped into \textit{Cl}$_{3,0}$,
\begin{equation}\label{HQW}
\sH_{QW}(\psi)=
\bk^2/2+\alpha_{\rm{R}}\e{12}\bk\psi\e{3},\quad\bk=k_x\e{1}+k_y\e{2}.
\end{equation}
This expression is a coordinate-free. The blades $\e{12}$ and
$\e{3}$ represent the QW plane and spin quantization axis which
may be arbitrary.

To find the eigens one has to solve the multivector equation
$\sH_{QW}(\psi)=E\psi$, where $E$ is the eigenenergy. Assuming
that the spinor is normalized $\psi\tilde{\psi}=1$, the equation
\eqref{HQW} can be rearranged into the rotor form
\begin{equation}\label{roteq12A}
\psi\e{3}\tilde{\psi}=\frac{k^2/2-E}{\alpha_R k^2}\bk\e{12},
\end{equation}
where $\bk\e{12}=\bk\d\e{12}$ is the vector and $k=|\bk|$ is $\bk$ norm. Since the right hand
side of \eqref{roteq12A} must be unit vector, its square must be
equal to $1$. This again results in a scalar equation (the quantization condition) the solution of which gives the spectrum (solid and dashed lines in
Fig.~\ref{f:DispEx}b),
\begin{equation}\label{eig12}
E_{\pm}=k^2/2\pm k\alpha_R.
\end{equation}
Insertion of eigenenergies $E_{\pm}$  back into \eqref{roteq12A} then
gives
\begin{equation}\label{roteq12B}
\psi_{\pm}\e{3}\tilde{\psi}_{\pm}=\pm(k_y\e{1}-k_x\e{2})/k\equiv\hat{\ba}_{\pm}.
\end{equation}
Since $\e{3}$ is perpendicular to, while $\hat{\ba}_{\pm}$ lies in
the quantum well plane the rotation angle is $\theta=\pm\pi/2$.
Then the solution of equation~\eqref{roteq12B} is the rotor
\begin{equation}
\psi_{\pm}=\exp\Big(\pm\frac{\e{3}\w\hat{\ba}_{\pm}}{|\e{3}\w\hat{\ba}_{\pm}|}\frac{\pi}{4}\Big).
\end{equation}
In  coordinate form the eigenspinors are
$\psi_{\pm}=(1\mp\cos\varphi\,\e{23}\mp\sin\varphi\,\e{31})\big/\sqrt{2}$.
The eigenspinors $\psi_{\pm}$ can be used to calculate the average spin
$\langle\mathbf{s}_{\pm}\rangle=\psi_{\pm}\e{3}\tilde{\psi}_{\pm}$,
\begin{equation}
\langle\mathbf{s}_{\pm}\rangle=\pm(\sin\varphi\,\e{1}-\cos\varphi\,\e{2}).
\end{equation}
Thus the spins of  SO split energy bands, as shown in Fig.~\ref{f:DispEx}b,
are parallel to confining planes and have opposite directions, and
at the same time (since $\bk\d\langle\mathbf{s}\rangle=0$) they
are perpendicular to the wave vector.

%%%%%%%%%%%%%%%%%%%%%%%%%%%%%%%%%%%%%%%%%%%%
\subsection{A pair of coupled two-level atoms}
In \textit{Cl}$_{3,0}$ algebra, a single two-level atom
Hamiltonian is $\hat{H}_1=\hbar\omega\hat{\sigma}_z/2$ which
 can be coupled to dipole moment
$\hat{\mathbf{d}}_1=\boldsymbol{\mu}\hat{\sigma}_x$, where
$\hbar\omega$ is the energy difference between the energy levels
and $\boldsymbol{\mu}$ is the vectorial coupling constant. Let us
take a pair of two-level atoms coupled to each other through
dipole-dipole interaction. This kind of interaction may produce
entanglement. Let the dipole coupling interaction energy be of the
form $\Gamma\hat{\sigma}^{(1)}\hat{\sigma}^{(2)}$, where $\Gamma$
is the coupling constant and the superscript indicates the first
and second two-level atom. Since the coupling energy is
proportional to $\boldsymbol{\mu}^{(1)}\d\boldsymbol{\mu}^{(2)}$,
the coupling constant  $\Gamma$ depends on angle $\theta$ between
the vectors $\boldsymbol{\mu}^{(1)}$ and $\boldsymbol{\mu}^{(2)}$.
The composite Hamiltonian describing two two-level atom then is
\begin{equation}\label{twoatoms}
\hat{H}_2=\frac{\hbar\omega}{2}\Big(\hat{\sigma}_z^{(1)}\otimes\hat{1}+\hat{1}\otimes\hat{\sigma}_z^{(2)}\Big)+
\Gamma\hat{\sigma}_x^{(1)}\otimes\hat{\sigma}_x^{(2)}.
\end{equation}
The eigenvalues of $\hat{H}_2$ form four energy levels,
$\pm\Gamma$ and $\pm\sqrt{\Gamma^2+\omega^2}$,  as shown in
Fig.~\ref{f:DispEx}c.

Since \textit{Cl}$_{3,0}$ algebra is too small for the four level
system, the \textit{Cl}$_{3,1}$ algebra will be used in the
following. Its matrix representation by $4\times 4$ complex
matrices is given in the Appendix~\ref{append}.  After mapping of
matrix equation~\eqref{twoatoms} onto \textit{Cl}$_{3,1}$ algebra
one finds the respective GA Hamiltonian ($\hbar=1$)
\begin{equation}\label{H2}
\sH_2(\psi)=\frac{\omega}{2}\e{34}\psi\e{34}-\Gamma\e{2}\psi\e{3}+\frac{\omega}{2}\e{3}\psi\e{3}.
\end{equation}
To find the eigens of the equation $\sH_2(\psi)=E\psi$ we will
utilize spatial inversion operation $\inv{\psi}=-\e{4}\psi\e{4}$
(see Appendix~\ref{append}). With the help of it the spinor can be
split into a sum of even $\inv{\psi}_{+}=\psi_{+}$ and odd
$\inv{\psi}_{-}=-\psi_{-}$ parts. Since $\e{34}\psi\e{34} =
\e{3}\bar{\psi} \e{3}$ and the eigenequation
$\sH_2(\psi)=E\psi$ and its inversion
$\inv{\sH}_2(\psi)=E\inv{\psi}$ give
$\big(\sH_2(\psi)+\inv{\sH}_2(\psi)\big)/2=E\psi_{+}$ and
$\big(\sH_2(\psi)-\inv{\sH}_2(\psi)\big)/2=E\psi_{-}$, we find
that the equations for even and odd states decouple into two independent equations,
\begin{align}
-\Gamma\e{2}\psi_{+}\e{3}&=E\psi_{+},\label{eq1}\\
\omega\e{3}\psi_{-}\e{3}-\Gamma\e{2}\psi_{-}\e{3}&=E\psi_{-}.\label{eq2}
\end{align}
Assuming that $\psi_{+}\tilde{\psi}_{+}=1$, the first
equation~\eqref{eq1} can be rearranged into rotor form
\begin{equation}
\psi_{+}\e{3}\tilde{\psi}_{+}=-\frac{E}{\Gamma}\e{2},
\end{equation}
from which we find the eigenvalues and respective rotor equation
\begin{equation}
E^{(1,2)}=\pm\Gamma,\quad \psi_{+}\e{3}\tilde{\psi}_{+}=\mp\e{2}.
\end{equation}
The solution of the latter gives the first pair of spinor-rotors
\begin{equation}
\psi_{+}^{(1,2)}=\ee^{\pm\e{23}\pi/4}=\big(1\pm\e{23}\big)/\sqrt{2}\,.
\end{equation}
The rotor $\psi_{+}^{(1)}$  gives the eigenenergy
$E_{+}^{(1)}=\langle\tilde{\psi}_{+}^{(1)}\sH(\psi_{+}^{(1)})\rangle=-\Gamma$
and rotates $\e{3}$ to $\e{2}$, while $\psi_{+}^{(2)}$ gives
$E_{+}^{(2)}=\langle\tilde{\psi}_{+}^{(2)}\sH(\psi_{+}^{(2)})\rangle=\Gamma$
and rotates  $\e{3}$ to $-\e{2}$.

In a similar fashion from the second eigenequation~\eqref{eq2} one
finds
\begin{equation}
E^{(3,4)}=\pm\sqrt{\omega^2+\Gamma^2},\quad
\psi_{-}\e{3}\tilde{\psi}_{-}=\pm\frac{\omega\e{3}-\Gamma\e{2}}{\sqrt{\omega^2+\Gamma^2}}\equiv\hat{\ba}.
\end{equation}
The solution of the second equation gives the second pair of
spinor-rotors
\begin{equation}\
\psi_{-}^{(3,4)}=\exp\Bigl({\pm\frac{\e{3}\w\hat{\ba}}{|\e{3}\w\hat{\ba}|}
\frac{\theta}{2}}\Bigr)=\cos\frac{\theta}{2}\pm\frac{\e{3}\w\hat{\ba}}{|\e{3}\w\hat{\ba}|}\sin\frac{\theta}{2},
\end{equation}
where $\cos\theta=\e{3}\d\hat{\ba}$,
$\cos(\theta/2)=2^{-1/2}\sqrt{1+\cos\theta}$,
$\sin(\theta/2)=2^{-1/2}\sqrt{1-\cos\theta}$.  In our case
$\cos\theta=\omega/\sqrt{\omega^2+\Gamma^2}$ and
$\frac{\e{3}\w\hat{\ba}}{|\e{3}\w\hat{\ba}|}=\e{23}$, therefore,
the rotation acts in $\e{23}$ plane.

%%%%%%%%%%%%%%%%%%%%%%%%%%%%%%%%%%%%%%
\subsection{Bilayer graphene\label{sec:4d}}
\begin{figure}
 \centering
 \begin{minipage}[c]{0.4\textwidth}
 \raggedright
\includegraphics[width=5.5cm]{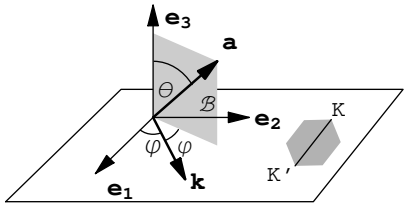}
 \end{minipage}%
 \begin{minipage}[c]{0.7\textwidth}
 \raggedleft
\caption{Basis vectors $\{\e{1},\e{2},\e{3}\}$, the wave vector
$\bk$, and the rotation plane
$\hat{\mathcal{B}}=\e{3}\w\hat{\ba}$. The vector $\e{3}$ is
perpendicular to bilayer graphene plane (square). The line that
connects $K$ and $K^{\prime}$ valleys in the Brillouin zone is
perpendicular to $\e{2}$.}\label{f:grapheneAxesab}
 \end{minipage}%
\end{figure}
In previous example the rotors $\psi_{+}$ and $\psi_{-}$ were
decoupled. Here we shall consider more complex case of bilayer
graphene (BG)  where the rotors are coupled. A unique feature of
the BG is its tunable energy band structure~\cite{Katsnelson12}.
In \textit{Cl}$_{3,1}$ the BG Hamiltonian function
reads~\cite{Dargys14c}
\begin{equation}\label{HBG}
\sH_{BG}(\psi)=\eta\bk\psi\Ie{3}-\frac{\gamma_1}{2}
\e{2}\big(\psi-\inv{\psi}\big)\e{3}+\eta U\e{3}\psi\e{3},
\end{equation}
where $\gamma_1$ is the interlayer coupling energy (constant),
$2U$ is the potential difference between the layers, and $\eta=+1$ is for
$K$-valley and $\eta=-1$ for $K^{\prime}$-valley, i.~e. the points in the momentum space where electron transport takes place. The presence of
two basis vectors in~\eqref{HBG} comes from the geometry. One of them represents pseudospin quantization axis which is aligned with $\e{3}$, the other being
$K-K^{\prime}$ axis which is perpendicular to $\e{2}$,
Fig.~\ref{f:grapheneAxesab}. Thus, the geometry again comes explicitly in GA formulation of
quantum problems.

To transform  the  eigenequation $\sH_{BG}(\psi)=E\psi$ into rotors
the spinor is again split into even and odd parts,
$\psi=\psi_{+}+\psi_{-}$, and then the  eigenequation is
decomposed into two coupled multivector equations for $\psi_{+}$
and $\psi_{-}$,
\begin{equation}\label{coupled}
\begin{split}
E\psi_{+}&=\eta\bk\psi_{-}\Ie{3}+\eta U\e{3}\psi_{+}\e{3},\\
E\psi_{-}&=\eta\bk\psi_{+}\Ie{3}+\eta
U\e{3}\psi_{-}\e{3}-\gamma_1\e{2}\psi_{-}\e{3}.
\end{split}
\end{equation}
The method of solution of the coupled system~\eqref{coupled} uses
the fact that $\psi_{+}$ and $I\psi_{-}$, where $I$ is the
pseudoscalar of \textit{Cl}$_{3,1}$, are rotors in 3D
Euclidean space. If $\psi_{-}$ is solved from the first
equation and inserted into the second equation of~\eqref{coupled}
then after some algebraic manipulations one can construct the following
rotor equation $\psi_{+}\e{3}\tilde{\psi}_{+}=\hat{\ba}$ for
$\psi_{+}$ (and analogous equation for $I\psi_{-}$). The
requirement that the final vector $\hat{\ba}$ in
Fig.~\ref{f:grapheneAxesab} after rotation will not change its
length is expressed by condition $\hat{\ba}^2= 1$ which
gives~\cite{Dargys14c}
\begin{equation}\label{conda}
\hat{\ba}^2=\frac{(E^2-k^2+U^2)^2+U^2\gamma_1^2}{E^2(4U^2+\gamma_1^2)}=1.
\end{equation}
This fourth order  polynomial equation yields BG spectrum shown in
Fig.~\ref{f:DispEx}d,
\begin{equation}\label{spectrum}
E^{(i)}=\pm\sqrt{k^2+U^2+\frac{\gamma_1^2}{2}\pm\frac{1}{2}\sqrt{\gamma_1^4+4k^2(4U^2+\gamma_1^2)}}\,.
\end{equation}
Knowing  $\hat{\ba}^{(i)}=\ba^{(i)}/|\ba^{(i)}|$ and $E^{(i)}$,
one can  construct the rotor for the $i$-th energy band. For
details and properties of Berry's phase calculated in terms of GA
the reader should refer to~\cite{Dargys14c}.

In conclusion, we have shown that presence of a specific direction
called the quantization axis in the Hamiltonian can be employed to
solve quantum problems in GA.  Since GA approach is
coordinate-free the formulas found in this way appear to be
compact and may be interpreted geometrically.

%%%%%%%%%%%%%%%%%%%%%%%%%%%%%%%%
\section{Appendix\label{append}}
\enlargethispage{8pt}
\textit{Matrix representation}. The basis vectors $\e{i}$ of
\textit{Cl}$_{3,1}$ algebra satisfy $\e{1}^2=\e{2}^2=\e{3}^2=1$
and $\e{4}^2=-1$. The following $4\times 4$ matrix representation
of $\e{i}$ was used in transforming Hilbert space Hamiltonians to
GA
\begin{equation}\label{e1e2}
\hat{e}_{1}=
\begin{bmatrix}
0&\hat{1}\\
\hat{1}&0\\
\end{bmatrix}
, \quad
\hat{e}_{2}=\ii
\begin{bmatrix}
0&-\hat{1}\\
\hat{1}&0\\
\end{bmatrix}
,\quad
\hat{e}_{3}=
\begin{bmatrix}
\hat{\sigma}_y&0\\
0&-\hat{\sigma}_y\\
\end{bmatrix}, \quad
\hat{e}_{4}=\ii
\begin{bmatrix}
\hat{\sigma}_z&0\\
0&-\hat{\sigma}_z\\
\end{bmatrix},
\end{equation}
where $\ii=\sqrt{-1}\,$. $\hat{1}$ is $2\times 2$ unit matrix and
$\hat{\sigma}_x$, $\hat{\sigma}_y$, $\hat{\sigma}_z$  are Pauli
matrices. Representations of higher blades can be found from
matrix products, for example,
$\e{12}\rightarrow\hat{e}_{12}=\hat{e}_{1}\hat{e}_{2}$.

\textit{Replacement rules} The following mapping is assumed
between the complex spinor $|\psi\rangle$ in a form of column and
GA spinor $\psi$,
\begin{equation}\label{psipsi}
 |\psi\rangle=
\begin{bmatrix}
\ \,a_0+\ii a_3\\
- b_3+\ii b_0\\
-b_2-\ii b_1\\
-a_1+\ii a_2
\end{bmatrix}
\longleftrightarrow\begin{cases}\psi=&a_0+a_1\e{23}-a_2\e{31}+a_3\e{12}\\
&\ -b_0I-b_1\e{14}+b_2\e{24}+b_3\e{34},
\end{cases}
\end{equation}
where $a_i$'s and $b_i$'s are scalars. Matrix representation of
basis elements \eqref{e1e2} and rule~\eqref{psipsi}  allow to construct the following replacement
rules between the action of matrices on column vectors and
\textit{Cl}$_{3,1}$ multivector functions,
\begin{equation}\label{repRulesAA}
\hat{e}_i|\psi\rangle\leftrightarrow\e{i}\psi\Ie{3},\quad
\hat{e}_{ij}|\psi\rangle\leftrightarrow\e{ij}\psi,\quad
\hat{I}\hat{e}_i|\psi\rangle\leftrightarrow\Ie{i}\psi\Ie{3},
\end{equation}
where $i,j=1,2,3,4$. Also, additional replacement rules may be
useful
\begin{equation}\label{repRulesA}
\begin{split}
&\ii|\psi\rangle\leftrightarrow I\psi\e{34},\quad
\langle\psi|\psi\rangle\leftrightarrow\langle\psi^{\dagger}\psi\rangle,\quad
\langle\varphi|\psi\rangle\leftrightarrow\langle\varphi^{\dagger}\psi\rangle-\langle\varphi^{\dagger}\psi\,\e{12}\rangle\e{12}.
\end{split}
\end{equation}
As always, in GA the angled brackets, for example $\langle
M\rangle$, indicate that only the scalar part of the
multivector~$M$ should be taken.

\textit{Spatial inversion, reversion and dagger operations}. The
spatial inversion  (denoted by overbar) changes signs of all
spatial vectors to opposite but  leaves ``time'' vector $\e{4}$
invariant: $\inv{\e{}}_i=-\e{i}$ for $i$=1,2,3 and
$\inv{\e{}}_4=\e{4}$. In \textit{Cl}$_{3,1}$  the inversion of
general multivector $M$ is defined by
\begin{equation}
\inv{M}=-\e{4}M\e{4}.
\end{equation}
Properties of inversion:  $\inv{M_1+M_2}=\inv{M}_1+\inv{M}_2$, and
$\inv{M_1M_2}=\inv{M}_1\,\,\inv{M}_2$. The spatial inversion
allows to split the general spinor~\eqref{psipsi} into even and
odd parts, $\psi=\psi_{+}+\psi_{-}$, which satisfy
$\inv{\psi}_{+}=\psi_{+}$ and $\inv{\psi}_{-}=-\psi_{-}$.

The reversion (denoted by tilde) changes the order of basis
vectors in the multivector. The dagger operation is a
combination of the reversion and spatial inversion
\begin{equation}\label{dagger}
\psi^{\dagger}=-\e{4}\tilde{\psi}\e{4}.
\end{equation}
If applied to a general bispinor it allows to find the square of
the module
\begin{equation}
\langle\psi^{\dagger}\psi\rangle=
a_0^2+a_1^2+a_2^2+a_3^2+b_0^2+b_1^2+b_2^2+b_3^2.
\end{equation}

\enlargethispage{15pt}

\begin{thebibliography}{10}

\bibitem{Hestenes66}
D.~Hestenes, \textit{Space-Time Algebra}. Gordon and Breach, New
York, 1966.

\bibitem{Snygg97}
J.~Snygg, \textit{ Clifford Algebra (A Computational Tool for
Physicists)}. Oxford University Press, NewYork, 1997.

\bibitem{Doran03}
C.~Doran and A.~Lasenby, \textit{Geometric Algebra for
Physicists}, Cambridge University Press, Cambridge, 2003.

\bibitem{Perwass09}
C.~Perwass, \textit{Geometric Algebra with Applications in
Engineering}. Springer-Verlag, Berlin Heidelberg, 2009.

\bibitem{Sprossig01}
W.~Spr{\"o}ssig, \textit{Eigenvalue problems in the framework of
{C}lifford analysis}. Adv. Appl. Clifford Algebras {\bf 11}(S2),
(2001) 301.

\bibitem{Hestenes99}
D.~Hestenes, \textit{New Foundations of Classical Mechanics}.
Kluwer Academic Publishers, The Netherlands, 1999.

\bibitem{Hestenes84}
D.~Hestenes and Sobczyk, \textit{Clifford Algebra to Geometric
Calculus (A Unified Language for Mathematics and Physics)}.
D.~Reidel Publishing Company, Dordrecht, 1984.

\bibitem{Katsnelson12}
M.~I. Katsnelson, \textit{Graphene: {C}arbon in Two Dimensions}.
Cambridge University Press, Cambridge, 2012.

\bibitem{Dargys13a}
A.~Dargys, \textit{Monolayer graphene and quantum flatland from a
view point of  geometric algebra}. Acta Phys. Pol. A {\bf 124},
(2013) 732.

\bibitem{Dargys14c}
A.~Dargys and A.Acus, \textit{Pseudospin, velocity and Berry phase
in a bilayer graphene}. arXiv:1410.2038 [cond-mat.mes-hall]  (2014).

\end{thebibliography}
\end{document}